\documentclass[pra,aps,twocolumn,amsmath,amssymb]{revtex4}

\usepackage{graphicx}
\usepackage{epsfig}
\usepackage{amsmath}
\usepackage{graphics}

\begin{document}

\title{Anomalous tunneling of collective excitations and effects of superflow in the polar phase of a spin-1 spinor Bose-Einstein condensate}

\author{Shohei Watabe$^{1,2}$, Yusuke Kato$^{3}$, and Yoji Ohashi$^{1,2}$} 
\affiliation{$^{1}$ Department of Physics, Keio University, 3-14-1 Hiyoshi, Kohoku-ku, Yokohama 223-8522, Japan
\\
$^{2}$ CREST(JST), 4-1-8 Honcho, Kawaguchi, Saitama 332-0012, Japan
\\
$^{3}$ Department of Basic Science, The University of Tokyo, Tokyo 153-8902, Japan}

\begin{abstract} 
We investigate tunneling properties of collective modes in the polar phase of a spin-1 spinor Bose-Einstein condensate. This spinor BEC state has two kinds of gapless modes, i.e., Bogoliubov mode and spin-wave. Within the framework of the mean-field theory at $T=0$, we show that these Goldstone modes exhibit the perfect transmission in the low-energy limit. Their anomalous tunneling behaviors still hold in the presence of superflow, except in the critical current state. In the critical current state, while the tunneling of Bogoliubov mode is accompanied by finite reflection, the spin-wave still exhibit the perfect transmission, unless the strengths of a spin-dependent and spin-independent interactions take the same value. 
%
We discuss the relation between perfect transmission of spin-wave and underlying superfluidity through a comparison of wavefunctions of spin-wave and the condensate. 

\end{abstract}

\maketitle

\section{Introduction} 

The perfect transmission through a potential barrier, which is frequently referred to as the anomalous tunneling phenomenon, is now widely recognized as a fundamental property of the Bogoliubov excitation in a Bose superfluid~\cite{Kovrizhin2001,Kagan2003,Danshita2006,FujitaMThesis,Watabe2008,Kato2008,Ohashi2008,Watabe2009RefleRefra,Tsuchiya2008,Takahashi2009JLTP,Takahashi2009,Tsuchiya2009,WatabeKato2009,WatabeKatoLett,WatabeKatoFull,WatabeKatoOhashiFull}. This phenomenon is quite different from the simple tunneling problem discussed in a one-particle quantum mechanics, where the transmission probability always vanishes in the low-energy limit. The anomalous tunneling is known to also occur in the presence of superflow, except in the critical supercurrent state, where the tunneling of Bogoliubov mode is accompanied by finite reflection.
\par
In Refs.~\cite{WatabeKato2009,WatabeKatoLett,WatabeKatoFull,WatabeKatoOhashiFull}, we have extended the previous work for the anomalous tunneling behavior of Bogoliubov mode to spin-wave excitations in the ferromagnetic phase of a spin-1 spinor Bose-Einstein condensate (BEC). We clarified that, not only the Bogoliubov mode, the gapless transverse spin-wave also exhibits the perfect transmission in the low-energy limit. In the presence of a finite superflow, the perfect transmission of the spin-wave occurs when the magnitude of the spin-wave momentum $k$ equals the supercurrent momentum $q$. On the other hand, the longitudinal spin-wave with a finite excitation gap does not show the anomalous tunneling behavior.
\par
In this paper, we further extend our previous studies~\cite{WatabeKato2009,WatabeKatoLett,WatabeKatoFull,WatabeKatoOhashiFull} to the polar phase of a spin-1 spinor BEC. Recently, two of the authors~\cite{WatabeKatoFull} clarified that the spin-wave in this phase also exhibits the perfect transmission in the absence of superflow. In this paper, we examine how this anomalous tunneling behavior is affected by supercurrent. Within the mean-field theory at $T=0$, we show that, in contrast to the case of gapless ferromagnetic spin-wave, the polar spin-wave always tunnel through a barrier without reflection {\it in the low-energy limit}. Although this property is the same as the anomalous tunneling behavior of Bogoliubov mode, we clarify that the perfect transmission of polar spin-wave occurs even in the critical current state, except for the case in which the spin-{\it dependent} interaction strength equals the spin-{\it independent} interaction strength. This critical behavior is quite different from the Bogoliubov mode, which is accompanied by finite reflection.
\par
This paper is organized as follows. In Sec.\ref{SecII}, we explain the outline of our formulation to deal with the polar state of a spin-1 spinor BEC. In Sec.\ref{SecIII}, we examine the tunneling properties of spin-wave excitations and effects of superflow in the polar phase. 

\begin{figure}[tbp]
\begin{center}
\includegraphics[width=8cm]{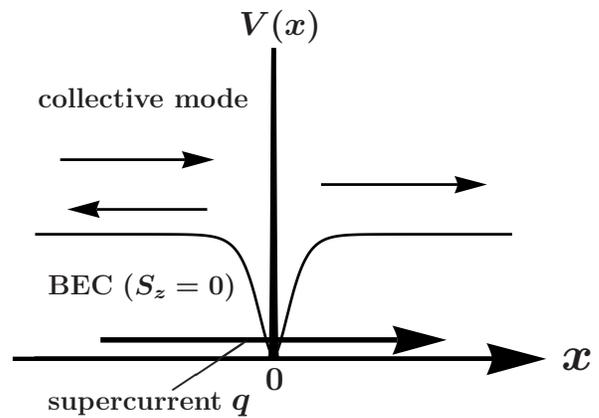}
\end{center}
\caption{Model one-dimensional tunneling system. The system is in the polar state ${\hat \Phi}(x,t)=(0,\Phi_0(x,t),0)$ with a finite superflow in the positive $x$-direction. We examine the transmission and refection of a collective mode injected from $x=-\infty$ by the potential barrier $V(x)$. 
}
\label{fig1}
\end{figure} 

\section{Mean-field theory for the polar state of spin-1 spinor BEC}\label{SecII}
\par
We consider the polar state of a spin-1 spinor BEC, and examine the tunneling of a collective mode through a potential barrier $V(x)$ put around $x=0$. Assuming the uniformity of the barrier in the $y$- and $z$-direction, we treat this problem as a one-dimensional system, described by the Lagrangian density~\cite{Ho1998,Ohmi1998,WatabeKatoOhashiFull},
\begin{align}
{\mathcal L}=
&
i\hbar {\hat \Phi}^\dagger (x,t)\partial_t{\hat \Phi}(x,t) 
\nonumber
\\
&
-
\Bigl[ 
{\hat \Phi}^\dagger(x,t)
\Bigl(
-\displaystyle{\hbar^2 \over 2m}\partial_x^2+V(x)
\Bigr)
{\hat \Phi}(x,t) 
\nonumber
\\
&
+
\displaystyle{c_{0} \over 2} \rho^{2} (x,t) 
+ 
\displaystyle{c_{1} \over 2} {\bf F}^{2} (x,t)
\Bigr].
\label{eq.1}
\end{align} 
Here, 
\begin{eqnarray}
{\hat \Phi}(x,t)=
\left(
\begin{array}{c}
\Phi_{+1}(x,t)\\
\Phi_0(x,t)\\
\Phi_{-1}(x,t)\\
\end{array}
\right)
\label{eq.2}
\end{eqnarray}
is a spinor Bose field with mass $m$, where $\Phi_{j}(x,t)$ $(j=\pm 1,0)$ describe three magnetic sublevels. $\rho(x,t)={\hat \Phi}^\dagger(x,t){\hat \Phi}(x,t)$ and ${\bf F}={\hat \Phi}^\dagger(x,t){\bf S}{\hat \Phi}(x,t)$, respectively, describe the particle density and spin density, where ${\bf S}=(S_x,S_y,S_z)$ are the $S=1$ spin matrices (where we take the spin quantization axis parallel to the $z$-axis). $c_0 = 4\pi\hbar^{2}(a_{0}+2a_{2})/(3m)$ and $c_1 = 4\pi\hbar^{2}(a_{2}-a_{0})/(3m)$, respectively, describe a spin-independent and spin-dependent interactions~\cite{Ho1998}, where $a_S$ is an $s$-wave scattering length in the channels of total spin $S=0,~2$. 
\par
In Ref.~\cite{WatabeKatoOhashiFull}, we have explained the detailed mean-field theory for the model system described by Eq. (\ref{eq.1}), to examine the tunneling problem in the spin-1 ferromagnetic spinor BEC. Since the previous formulation is also applicable to the present polar state, we only present the outline of our formulation here. For more details, we refer to Ref.~\cite{WatabeKatoOhashiFull}. 
\par
The time-dependent Gross-Pitaevskii equation for the condensate wavefunction ${\hat \Phi}(x,t)$ is given by
\begin{eqnarray}
i\hbar
{\partial {\hat \Phi}(x,t) \over \partial t}=
\left(
\begin{array}{ccc}
h_{+}(x,t) & \displaystyle{c_1 \over \sqrt{2}} F_-& 0\\
\displaystyle{c_1 \over \sqrt{2}}F_+ & h(x,t) & \displaystyle{c_1 \over \sqrt{2}}F_-\\
0 & \displaystyle{c_1 \over \sqrt{2}}F_+ & h_{-}(x,t) \\
\end{array}
\right)
{\hat \Phi}(x,t),
\label{eq.3}
\end{eqnarray}
where $h(x,t)=- \hbar^{2}\partial_x^2/(2m) + V(x) + c_{0} \rho(x,t)$, $h_{\pm} (x,t) \equiv h(x,t) \pm c_1F_z$, and $F_\pm=F_x\pm iF_y$. As usual, we study the tunneling problem in the stationary state, as schematically shown in Fig.\ref{fig1}. In this case, setting ${\hat \Phi}(x,t)=e^{-i\mu t/\hbar}{\hat \Phi}(x)$, we obtain the GP equation for the spatial part ${\hat \Phi}(x)$ as
\begin{eqnarray}
\left(
\begin{array}{ccc}
h_{+}(x) & \displaystyle{c_1 \over \sqrt{2}} F_-& 0\\
\displaystyle{c_1 \over \sqrt{2}}F_+ & h(x) & \displaystyle{c_1 \over \sqrt{2}}F_-\\
0 & \displaystyle{c_1 \over \sqrt{2}}F_+ & h_{-}(x) \\
\end{array}
\right)
{\hat \Phi}(x)=0.
\label{eq.4}
\end{eqnarray}
Here, $h(x)=- \hbar^{2}\partial_x^2/(2m) -\mu + V(x) + c_{0} \rho(x)$ and $h_{\pm}(x)=h(x) \pm c_1F_z$, 
where $\rho(x)={\hat \Phi}^\dagger(x){\hat \Phi}(x)$ is the condensate density. 
\par
\begin{figure}[tbp]
\begin{center}
\includegraphics[width=6cm]{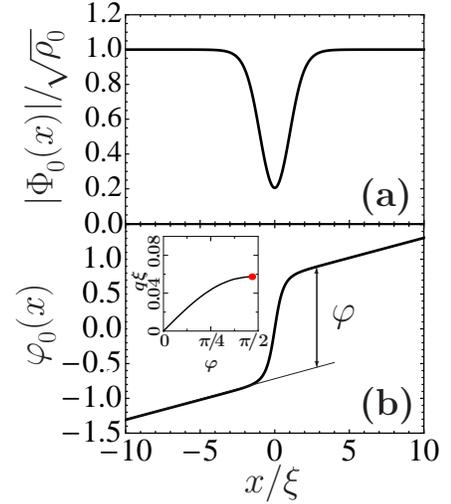}
\end{center}
\caption{(Color online) Calculated condensate wavefunction $\Phi_{0}(x)=|\Phi_0(x)|e^{i\varphi_0(x)}$ around the potential barrier $V(x) = 2 c_{0} \rho_{0} \exp{[ - (x/\xi )^{2} ]}$ in the critical current state. (a) Amplitude $|\Phi_0(x)|$. (b) Phase $\varphi_0(x)$. In panel (b), we set $\varphi_0(0)=0$. The inset shows the relation between the phase difference $\varphi$ (which is given in panel (b)~\cite{note100}) and supercurrent momentum $q$, where $\xi=\hbar/\sqrt{mc_0\rho_0}$ is the healing length. The red point at $q=0.0574\xi^{-1}$ ($\equiv q_{\rm c}$) represents the critical current state.
}
\label{fig2}
\end{figure}

The polar state becomes the most stable state when the spin-dependent interaction is antiferromagnetic ($c_1>0$). In this case, the condensate wavefunction ${\hat \Phi}(x)$ has the form~\cite{NoteZero},
\begin{eqnarray}
{\hat \Phi}(x)=
\left(
\begin{array}{c}
0\\
\Phi_0(x)\\
0\\
\end{array}
\right),
\label{eq.5}
\end{eqnarray}
which leads to the vanishing spin density ${\bf F}={\hat \Phi}^\dagger(x){\bf S}{\hat \Phi}(x)$~\cite{Polar}. The GP equation for the non-vanishing $\Phi_0(x)$-component is then given by
\begin{equation}
h(x)\Phi_0(x)=0.
\label{eq.6}
\end{equation}
In the supercurrent state, since the superfluid properties far away from the barrier are the same as those in the uniform system, one finds $\mu=c_0 \rho_0+\hbar^2q^2/(2m)$ (where $q$ is the supercurrent momentum and $\rho_0$ is the condensate density at $x=\pm\infty$). We show in Fig. \ref{fig2} the calculated spatial variation of the condensate wavefunction $\Phi_0(x)$ in the supercurrent state.
\par

\par
A collective mode in the polar phase can be conveniently calculated by considering fluctuations of the condensate wavefunction around the mean-field value in Eq. (\ref{eq.5}). In the time-dependent GP equation (\ref{eq.3}), setting ${\hat \Phi}(x,t)=\exp{(-i\mu t/\hbar)} ( {\hat \Phi(x)}+{\hat \phi}(x,t))$ (where ${\hat \phi}=(\phi_{+1},\phi_0,\phi_{-1})$ describes fluctuations around the mean-field value $\Phi_0(x)$), and retaining terms to $O({\hat \phi}(x))$, we obtain
\begin{align}
i\hbar \frac{\partial {\phi}_{0} (x, t) }{\partial t} 
=  & 
h(x,c_{0})
{\phi}_{0} (x) 
+ 
c_{0} \Phi_{0}^{2} (x) {\phi}_{0}^{*} (x), 
\label{eq.7}
\\
i\hbar \frac{\partial {\phi}_{\pm 1} (x, t) }{\partial t} 
= &
h(x,c_{1})
{\phi}_{\pm 1} (x) 
+ 
c_{1} \Phi_{0}^{2} (x) {\phi}_{\mp 1}^{*} (x), 
\label{eq.8}
\end{align}
where $h(x,c_{0,1}) \equiv h(x) + c_{0,1} |\Phi_{0} (x) |^{2} $. 
Since Eq.(\ref{eq.7}) for ${\phi}_{0}$ is decoupled from Eq.(\ref{eq.8}) for ${\phi}_{\pm 1}$, one may separately treat them. The former equation gives the Bogoliubov excitation. Indeed, setting ${\phi}_{0} (x, t) = u (x) e^{-iEt/\hbar} - v^{*} (x) e^{+iEt/\hbar}$, we obtain the ordinary Bogoliubov equation,
\begin{eqnarray}
E 
\begin{pmatrix}
u (x) \\ v (x) 
\end{pmatrix}
= 
\begin{pmatrix}
{h} (x, c_{0}) & - c_{0} \Phi_{0}^{2} (x) 
\\
c_{0} \Phi_{0}^{*2} (x) & - {h} (x,c_{0}) 
\end{pmatrix} 
\begin{pmatrix}
u (x) \\ v (x)
\end{pmatrix}. 
\label{Bogo}
\end{eqnarray} 
On the other hand, one may interpret $\phi_{\pm 1}(x,t)$ as fluctuations of $S_z=\pm 1$-components around $\Phi_{\pm 1}=0$, so that they give spin-wave excitations. Substituting ${\phi}_{\pm 1} (x,t) = \phi_{\pm 1} (x) e^{\mp i E t/\hbar}$ into Eq. (\ref{eq.8}), one obtains the ``spin-wave equation,"
\begin{align}
E 
\begin{pmatrix}
\phi_{+1} (x) \\
\phi_{-1}^{*} (x) 
\end{pmatrix}
=
\begin{pmatrix}
{h} (x, c_{1}) & c_{1} \Phi_{0}^{2} (x) \\
- c_{1} \Phi_{0}^{*2} (x) & - {h} (x, c_{1}) 
\end{pmatrix} 
\begin{pmatrix}
\phi_{+1} (x) \\
\phi_{-1}^{*} (x) 
\end{pmatrix}. 
\label{BogoPM1}
\end{align}
When $(\phi_{+1}(x),\phi_{-1}(x))$ is a solution of Eq. (\ref{BogoPM1}), $(\phi_{-1}(x)e^{iEt/\hbar},\phi_{+1}(x)e^{-iEt/\hbar})$ also satisfies Eq. (\ref{eq.8}), which means that the spin-wave excitations in the polar state are doubly degenerate.
\par
The set of GP equation (\ref{eq.6}) and the Bogoliubov equation (\ref{Bogo}) is the same as that in the scalar BEC. Thus, the Bogoliubov excitation in the polar phase of a spin-1 spinor BEC has the same tunneling properties as that in the scalar BEC. That is, the perfect transmission occurs in the long-wavelength limit, and this anomalous tunneling phenomenon still holds in the presence of a finite superflow, except in the critical current state. In the critical current state, the transmission of low-energy Bogoliubov mode is accompanied by finite reflection.
\par
The spin-wave equation (\ref{BogoPM1}) reduces to the Bogoliubov equation (\ref{Bogo}) when $c_1=c_0$~\cite{note}. When this relation is satisfied, spin-wave excitations are found to also exhibit the anomalous tunneling behavior as in the case of Bogoliubov mode. However, one cannot use this analogy when $c_1\ne c_0$. The goal of the next section is to clarify what happens in this case.
\par
In considering the supercurrent state, we need to choose the supercurrent momentum $q$ so that the Landau instability will not occur. In the uniform system, Eqs.(\ref{Bogo}) and (\ref{BogoPM1}), respectively, give the Bogoliubov and spin-wave excitation spectra as,
\begin{align}
E = {\hbar^2 q k \over m}
+ \sqrt{\frac{\hbar^2k^{2}}{2m} \left ( \frac{\hbar^2 k^{2}}{2m} + 2c_0\rho_0 \right )}, 
\label{eq;energy1}
\end{align} 
\begin{align}
E = {\hbar^2 q k \over m}
+ \sqrt{\frac{\hbar^2 k^{2}}{2m} \left ( \frac{\hbar^2 k^{2}}{2m} + 2c_1\rho_0 \right )}. 
\label{eq;energy2}
\end{align}
Thus, the stability condition for the supercurrent state, which is guaranteed when Eqs.(\ref{eq;energy1}) and (\ref{eq;energy2}) are always positive, is obtained as
\begin{equation}
|q|\xi\le {\rm Min}[1,\sqrt{c_1/c_0}].
\label{stability}
\end{equation}
\par
For later convenience, we introduce the dimensionless variables, $\bar{E} \equiv E/(c_{0}\rho_{0})$, $\bar{x} \equiv x/\xi$, $\bar{k} = k\xi$, $\bar{q} = q\xi$, $\bar{V} \equiv V  / (c_{0}\rho_{0})$, $\bar{\Phi}_{i} \equiv \Phi_{i}/\sqrt{\rho_{0}}$, $\overline{{\phi}}_{i} \equiv {\phi}_{i}/\sqrt{\rho_{0}}$, $\bar{t} \equiv t c_{0} \rho_{0} / \hbar$ and $\bar{c} \equiv c_{1}/c_{0}$, where $\xi=\hbar/\sqrt{mc_0\rho_0}$ is the healing length. Using these, we examine the low-energy tunneling properties of the polar spin-wave in the next section. For simplicity, we omit the bar in what follows.
\par


\section{Anomalous Tunneling of Spin-wave Modes in the current carrying polar phase}\label{SecIII}
\par
We numerically solve the spin-wave equation (\ref{BogoPM1}), together with the GP equation (\ref{eq.6}), in the presence of a finite superflow with the momentum $q$. As shown in Fig.\ref{fig1}, we consider the case that an incident spin-wave with the energy $E$ is injected from $x\ll -1$, and tunneling through the barrier $V(x)$ around $x=0$. The wavefunction $(\phi_{+1}(x),\phi_{-1}(x))$ obeys the boundary conditions,
\begin{align}
\begin{pmatrix}
\phi_{+1} \\ \phi_{-1}^{*} 
\end{pmatrix}
= &
\begin{pmatrix}
\alpha_{k_{1}} e^{+iqx} \\ \beta_{k_{1}} e^{-iqx}
\end{pmatrix}
e^{i k_{1} x} 
+ 
R
\begin{pmatrix}
\alpha_{k_{2}} e^{+iqx} \\ \beta_{k_{2}} e^{-iqx} 
\end{pmatrix}
e^{i k_{2} x} 
\nonumber
\\
&
+
A
\begin{pmatrix}
\alpha_{k_{3}} e^{+iqx} \\ \beta_{k_{3}} e^{-iqx}
\end{pmatrix}
e^{i k_{3} x} ( x \ll -1 ), 
\label{eqBoundary1}
\\
\begin{pmatrix}
\phi_{+1} \\ \phi_{-1}^{*} 
\end{pmatrix}
= &
T
\begin{pmatrix}
\alpha_{k_{1}} e^{+iqx} \\ \beta_{k_{1}}  e^{-iqx}
\end{pmatrix}
e^{i k_{1} x} 
\nonumber
\\
&
+
B 
\begin{pmatrix}
\alpha_{k_{4}} e^{+iqx} \\ \beta_{k_{4}}  e^{-iqx}
\end{pmatrix}
e^{i k_{4} x} (x \gg +1).
\label{eqBoundary2}
\end{align}
Here, each term in Eqs. ({\ref{eqBoundary1}) and (\ref{eqBoundary2}) is an independent solution of Eq. (\ref{BogoPM1}) in the absence of barrier $V(x)$.  That is, the coefficients $\alpha_k$ and $\beta_k$ are given by
\begin{eqnarray}
\left(
\begin{array}{c}
\alpha_{k}\\
\beta_{k}
\end{array}
\right) 
= & \displaystyle{\frac{1}{\sqrt{c^{2} -|E - c - qk - k^{2}/2|^2}}} 
\\ 
& \times
\left(
\begin{array}{c}
{c}\\
{E-c-qk-k^2/2}
\end{array}
\right), 
\label{eq.ab}
\end{eqnarray}
where the normalization condition is imposed as $|\phi_{+1} |^{2} - |\phi_{-1} |^{2} = 1$ following the Bogoliubov mode~\cite{FetterWalecka}. 
The momenta $k_j$ ($j=1\sim 4$) are the solutions of 
\begin{align}
k^{4} + 4(c_{} - q^{2}) k^{2} + 8 k q E - 4 E^{2} = 0. 
\label{Equatiok}
\end{align} 
Equation (\ref{Equatiok}) has two propagating solutions ($j=1,2$), as well as two damping solutions ($j=3,4$). In the low energy region, they are given by
\begin{eqnarray}
\left(
\begin{array}{c}
k_{1}\\
k_{2}
\end{array}
\right)
\simeq 
\left(
\begin{array}{c}
\displaystyle
\frac{E}{\sqrt{c} + q}\\
\displaystyle
\frac{-E}{\sqrt{c} - q}
\end{array}
\right), 
\label{k1k2}
\end{eqnarray}
\begin{eqnarray}
\left(
\begin{array}{c}
k_{3}\\
k_{4}
\end{array}
\right)
\simeq 
\left(
\begin{array}{c}
\displaystyle
\frac{q}{c-q^{2}} E- 2 i \sqrt{c -q^{2}} \\
\displaystyle
\frac{q}{c-q^{2}} E+ 2 i \sqrt{c -q^{2}}  
\end{array}
\right). 
\label{k3k4}
\end{eqnarray}
\par
Once the coefficients $(T,R,A,B)$ are determined, the transmission probability can be calculated in the same manner as in the ordinary tunneling problem. From Eq. (\ref{eq.8}), we find that the generalized density $n_{\rm s}(x,t)=|\phi_{+1}(x,t)|^2-|\phi_{-1}(x,t)|^2$ satisfies the continuity equation $\partial_t n_{\rm s}+\partial_x J_{\rm s}=0$. Thus, in the stationary state, the flux density
\begin{equation}
J_{\rm s}={1 \over m}{\rm Im}[\phi_{+1}^*\partial_x\phi_{+1}+\phi_{-1}\partial_x\phi_{-1}^*]
\label{eq.J2}
\end{equation}
is a conserving quantity. Using this, one may conveniently introduce the transmission probability $\tau$, as well as the reflection probability $r$, as 
\begin{eqnarray} 
\tau = & \Bigl|{J_{\rm s}^{\rm t} \over J_{\rm s}^{\rm i}}\Bigr|= & |T|^2,
\\ 
r = &
\Bigl|{J_{\rm s}^{\rm r} \over J_{\rm s}^{\rm i}}\Bigr| = &
\Bigl|\frac{w(k_{2})}{w(k_{1})}\Bigr||R|^{2}, 
\end{eqnarray}
where $w(k)=k[|\alpha_k|^2+|\beta_k|^2]+q[|\alpha_k|^2-|\beta_k|^2]$. $J_{\rm s}^{\rm i}$ and $J_{\rm s}^{\rm r}$ describe the flux densities of incident and reflected waves, respectively, that are calculated from the first and second terms in the RHS of Eq.(\ref{eqBoundary1}), respectively. The flux density $J_{\rm s}^{\rm t}$ of the transmission wave is obtained from the first term in the RHS of Eq. (\ref{eqBoundary2}). 
\par
\begin{figure}[tbp]
\begin{center}
\includegraphics[width=8cm]{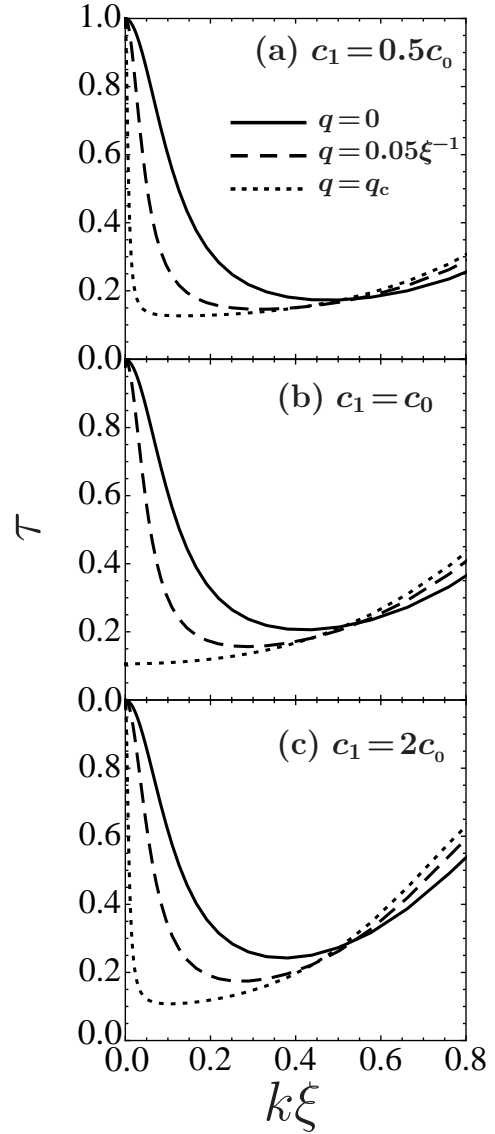}
\end{center}
\caption{Transmission probability $\tau$ of the spin-wave in the polar phase, as a function of incident momentum $k$. We take $V_0=2c_0\rho_0$ and $\lambda/\xi=1$, that give the same potential barrier as that used in Fig.~\ref{fig2}. 
}
\label{fig3}
\end{figure} 
\par

Figure~\ref{fig3} shows the transmission probability $\tau$ of the spin-wave through the potential barrier,
\begin{equation}
V(x) = V_0 e^{-(x/\lambda)^{2}}.
\label{eq.pot}
\end{equation}
In this figure, we find that the anomalous tunneling (perfect transmission) occurs in the supercurrent state of the polar phase, irrespective of the value of $c=c_1/c_0$. The perfect transmission can be seen even in the critical current state, except when $c=1$, which is quite different from the case of Bogoliubov mode, where the tunneling is accompanied by finite reflection ($r>0$). 
\par
We briefly note that the tunneling properties of the spin-wave mode shown in Fig.\ref{fig3} are quite different from the case of {\it ferromagnetic} spinor BEC~\cite{WatabeKatoOhashiFull}. In the latter case, the gapless transverse spin-wave mode exhibits the perfect transmission in the low-momentum limit only in the absence of supercurrent. In the supercurrent state, the perfect transmission occurs 
at the momentum whose magnitude equals that of the supercurrent momentum. 
The other spin-wave mode with a finite excitation gap does not show the anomalous tunneling behavior.
\par
As mentioned previously, the spin-wave equation (\ref{BogoPM1}) reduces to the ordinary Bogoliubov equation when $c=c_1/c_0=1$. 
Thus, the absence of the perfect transmission in the critical current state shown in Fig.\ref{fig3} (b) is due to the same mechanism as that in the case of Bogoliubov mode. In the latter case, in addition to the zero-energy phase mode solution given by
\begin{eqnarray}
\left(
\begin{array}{c}
u(x)\\
v(x)
\end{array}
\right)
=
\left(
\begin{array}{c}
\Phi_{0} (x)\\
\Phi_{0}^{*} (x)
\end{array}
\right),
\label{eq.zero1}
\end{eqnarray}
the Bogoliubov equation (\ref{Bogo}) has the other zero-energy solution in the critical current state~\cite{Takahashi2009},
\begin{eqnarray}
\left(
\begin{array}{c}
u(x)\\
v(x)
\end{array}
\right)
=
\left(
\begin{array}{c}
\partial_{\varphi} \Phi_{0} (x)\\
- \partial_{\varphi}\Phi_{0}^{*} (x)
\end{array}
\right),
\label{eq.an}
\end{eqnarray}
where $\varphi$ is the phase difference of the condensate wavefunction $\Phi_0(x)$ between $x=\pm\infty$~\cite{note100}. This solution physically describes density fluctuations near the barrier, leading to the suppression of perfect transmission~\cite{Takahashi2009}. Similarly, Eq. (\ref{BogoPM1}) always has the zero-energy solution,
\begin{eqnarray}
\left(
\begin{array}{c}
\phi_{+1}(x)\\
\phi_{-1}(x)
\end{array}
\right)
=
\left(
\begin{array}{c}
\Phi_{0} (x)\\
-\Phi_{0}^{*} (x)
\end{array}
\right),
\label{eq.zero3}
\end{eqnarray}
which describes the Goldstone mode associated with the broken spin rotational symmetry in the polar phase. In addition to this, one also has the other zero-energy solution in the critical current state, given by
\begin{eqnarray}
\left(
\begin{array}{c}
\phi_{+1}(x)\\
\phi_{-1}(x)
\end{array}
\right)
=
\left(
\begin{array}{c}
\partial_{\varphi} \Phi_{0} (x)\\
\partial_{\varphi}\Phi_{0}^{*} (x)
\end{array}
\right).
\label{eq.zero4}
\end{eqnarray}
While spin fluctuations described by $\delta F_{\pm}=\sqrt{2} (\Phi_{0} \phi_{\pm1}^{*} + \Phi_{0}^{*} \phi_{\mp1})$ vanishes in the case of Goldstone mode in Eq.(\ref{eq.zero3}), Eq.(\ref{eq.zero4}) gives $\delta F_\pm\ne 0$. Thus, in contrast to the case of Bogoliubov mode (where density fluctuations destroy the perfect transmission in the critical current state), the absence of the anomalous tunneling of the spin-wave when $c=1$ is due to the emergence of spin fluctuations.
\par
We note that Eq. (\ref{eq.zero4}) does not satisfy Eq. (\ref{BogoPM1}) unless $c=1$. Because of this, when $c\ne 1$, the spin-wave can tunnel through the barrier without being disturbed by spin fluctuations, even in the critical current state, as shown in Figs.~\ref{fig3} (a) and (c).
\par
We also note that, as pointed out in Ref.~\cite{KatoWatabe2010}, the emergence of the zero-energy density mode in Eq. (\ref{eq.an}) is deeply related to the instability of the superfluid state in the case of scalar BEC. In the polar state of a spin-1 spinor BEC, while both low-energy density fluctuations and spin fluctuations exist when $c=1$, the former fluctuations only appear when $c\ne 1$ in the critical current state. This implies that the mechanism of supercurrent instability above the critical current state might be somehow different between the cases of $c=1$ and $c\ne 1$.
\par
Equation (\ref{eq.zero3}) indicates that the general property of the zero-energy Bogoliubov mode, 
that each component of the wavefunction $(u(x),v(x))$ coincides with the condensate wavefunction $\Phi_0(x)$~\cite{Kato2008} (See Eq.(\ref{eq.zero1})), 
also holds in the spin-wave. In this regard, it has been also pointed out in the scalar BEC that such a coincidence also exists, when the Bogoliubov excitation has a small but finite momentum ($E\propto p>0$)~\cite{Ohashi2008}. 
To see if this property, which we call ``supercurrent behavior", is also applicable to the spin-wave, we consider the tunneling problem for a $\delta$-functional potential barrier $V(x)=V_0\delta(x)$. 
In this simple model, solving the GP equation (\ref{eq.6}) analytically, one obtains the condensate wavefunction in the supercurrent state with momentum $q$ as~\cite{Danshita2006,Ohashi2008}, 
\begin{align}
\Phi_{0} (x) = & e^{i (q x + \theta_{q} ) } 
[{\gamma}_{q}^{<} (x) + i q ] 
\nonumber \\
\equiv & \Psi_{<} (x,q) (x < 0), 
\label{DeltaPhi01}
\\
\Phi_{0} (x) = & e^{i (q x - \theta_{q} ) } 
[{\gamma}_{q}^{>} (x) - i q ] 
\nonumber \\
\equiv & \Psi_{>} (x,q) (x \geq 0), 
\label{DeltaPhi02}
\end{align}
where 
${\gamma}_{q}^{<} (x)$ is $\gamma_{q} (x)$ for $x < 0$,  
${\gamma}_{q}^{>} (x)$ is $\gamma_{q} (x)$ for $x \geq 0$, 
and $\exp{(i \theta_{q})} =[{\gamma}_{q} (0) - i q] / \sqrt{ {\gamma}_{q}^{2}(0) + q^{2} }$. 
Here, ${\gamma}_{q} (x) \equiv \sqrt{1 - q^{2}} \tanh {(\sqrt{1-q^{2}} (|x| + x_{0}) )}$, and 
$x_{0}$ is determined so as to satisfy $\partial_{x} \Phi_{0} (+0) - \partial_{x} \Phi_{0} (-0) = 2 V_{0} \Phi_{0} (0)$~\cite{Danshita2006}. 
In the low momentum region ($|p|,|q|\ll 1$), 
one obtains the following form:
\begin{eqnarray}
\left(
\begin{array}{c}
\phi_{+1} \\
\phi^*_{-1}
\end{array}
\right)
\simeq
\left(
\begin{array}{c}
\Lambda_+\Psi_{<}^{}(X_{-},p+q)\\
\Lambda_- \Psi_{<}^{} (X_{+},p-q)
\end{array}
\right)
, (x < 0), 
\label{eq.super}
\\
\left(
\begin{array}{c}
\phi_{+1} \\
\phi^*_{-1}
\end{array}
\right)
\simeq
\left(
\begin{array}{c}
\Lambda_+ \Psi_{>}(X_{+},p+q)\\
\Lambda_-  \Psi_{>}^{} (X_{-},p-q)
\end{array}
\right)
, (x \geq 0), 
\label{eq.super2}
\end{eqnarray}
with the accuracy of $O(|p|)$, and $O(|q|)$. Here $\Lambda_\pm$ are unimportant numerical factors which depend on $p$ and $q$; $X_{\pm}$ are functions of $x$. 
We note that the relations (\ref{eq.super}) and (\ref{eq.super2}) cannot be extended to $q=q_{\rm c}$ when $c = 1$. 
If (\ref{eq.super}) and (\ref{eq.super2}) held for $q=q_{\rm c}$ when $c = 1$, 
it would lead to partial transmission of spin-wave at low-energy limit. 

``Supercurrent behavior" was first proposed as an explanation of perfect transmission of the Bogoliubov mode through the potential barrier~\cite{Ohashi2008}. 
This property was successfully proved in the transverse spin-wave of the ferromagnetic spinor BEC for an arbitrary shape of the barrier~\cite{WatabeKatoOhashiFull}. 
The Bogoliubov mode and the spin-wave in the present polar phase certainly show ``supercurrent behavior" in the presence of the $\delta$-function potential barrier, but it is fair to say that the applicability of ``supercurrent behavior"  as an explanation of perfect transmission of these two modes through an arbitrary shape of the barrier is currently restrictive compared with the transverse spin-wave in the ferromagnetic spin-1 BEC.
\par

\section{Summary}
To summarize, we have investigated tunneling properties of collective excitations in the polar phase of spin-1 spinor BEC. We showed that the low-energy spin-wave can tunnel through a barrier without reflection, even in the presence of a finite superflow. In the critical current state, tunneling properties depend on the ratio $c=c_1/c_0$, where $c_0$ and $c_1$ are the strengths of a spin-independent and spin-dependent interactions, respectively. When $c=1$, the perfect transmission does not occur in the critical current state. Otherwise, the anomalous tunneling behavior continues to exist in the critical current state. 
\par
In the case of Bogoliubov mode, the anomalous tunneling phenomenon does not occur in the critical current state. In the critical current state, a density fluctuation mode appears around the barrier, which suppresses the perfect transmission of Bogoliubov mode. In the spin-wave case with $c=1$, spin fluctuations appear around the barrier in the critical current state, leading to the breakdown of the perfect transmission of the spin-wave. When $c\ne 1$, such spin fluctuations do not appear, so that one obtains the anomalous tunneling phenomenon of the spin-wave in the critical current state.
\par
We note that the anomalous tunneling behavior of the spin-wave mode in the polar phase is somehow different from the case of ferromagnetic spinor state. In the latter case, the gapless transverse spin-wave 
excitation exhibits the perfect transmission, not in the low-momentum limit, but in the case when the magnitude of the spin-wave momentum $k$ equals that of supercurrent momentum $q$. The longitudinal spin-wave with a finite excitation gap does not show the anomalous tunneling phenomenon. In this case, the perfect reflection occurs 
in the low-momentum limit, as in the case of the ordinary tunneling of a quantum mechanical particle. 
\par
However, when the perfect transmission occurs, the wavefunctions of Bogoliubov mode, polar spin-wave, and ferromagnetic transverse spin-wave, all have the same form as the condensate wavefunction. 
Through a series of papers and the present work, 
we have shown that anomalous tunneling is a common property of all gapless collective modes in scalar and the spin-1 spinor BECs.
\par

\acknowledgements
This work was supported by Grant-in-Aid for Scientific Research 
(Grant No. 20500044, 21540352, 22540412) from JSPS, Japan.


\appendix
\section{Derivation of Eqs. (\ref{eq.super}) and (\ref{eq.super2})}
\par
Expanding the condensate wavefunction in Eqs. (\ref{DeltaPhi01}) and (\ref{DeltaPhi02}) to $O(q)$, we obtain
\begin{align}
\Psi_{<} (x,q) \simeq  & e^{iqx} \left [ \gamma^{<} (x) -i q \frac{\gamma^{<} (x) - \gamma_{} (0) }{\gamma_{} (0) } \right ] (x<0), 
\label{ApproxDeltaPhi01}
\\
\Psi_{>} (x,q) \simeq & e^{iqx} \left [ \gamma^{>} (x) +  i q \frac{\gamma^{>} (x) - \gamma_{} (0) }{\gamma_{} (0) } \right ] (x \geq 0), 
\label{ApproxDeltaPhi02}
\end{align} 
where $\gamma^{<}(x)$ is $\gamma (x)$ for $x < 0$, 
$\gamma^{>}(x)$ is $\gamma (x)$ for $x \geq 0$. 
Here, $\gamma (x) \equiv \tanh(|x| + x_{0})$. 
In this limiting case, the boundary condition at $x=0$ ($\partial_x\Phi_0(+0)-\partial_x\Phi_0(-0)=2V_0\Phi_0(0)$) gives $V_0=[1-\gamma^2(0)]/\gamma(0)$. The goal of this appendix is to show that the components $\phi_{\pm 1}(x)$ in the wavefunction of the spin-wave with $|p|\ll 1$ reduces to 
the forms (\ref{ApproxDeltaPhi01}) and (\ref{ApproxDeltaPhi02}). 
\par
We introduce the functions $S(x)$ and $G(x)$, given by
\begin{align}
S (x) \equiv & \phi_{+1} (x) 
e^{- i \varphi_0 (x)} - \phi_{-1}^{*} (x) e^{+ i \varphi_0 (x)},  
\label{defS}
\\ 
G (x) \equiv & \phi_{+1} (x) 
e^{- i \varphi_0 (x)} + \phi_{-1}^{*} (x) e^{+ i \varphi_0 (x)}, 
\label{defG}
\end{align} 
where $\varphi_0(x)$ is the phase of the condensate wavefunction $\Phi_0(x)=|\Phi_0(x)|e^{i\varphi_0(x)}$. 
In the case of $\delta$-functional barrier~\cite{Danshita2006}, we obtain 
$\varphi_0(x)=qx+{\rm sgn}(x)[ \tan^{-1}(\gamma_q(x)/q) - \tan^{-1}(\gamma_q(0)/q)]$. Equations for $S(x)$ and $G(x)$ are obtained from Eq.(\ref{BogoPM1}) as
\begin{align}
EG(x) = & H (x) S(x) 
- i q 
\frac{\left \{ |\Phi_0(x)|, G(x) \right \}}{|\Phi_0(x)|^{3}}, 
\label{eqSGfull1}
\\
ES(x) = & \left [ H (x) +  2 c_{} |\Phi_0(x)|^{2} \right ] G(x) 
- i q \frac{ \left \{ |\Phi_0(x)|, S(x) \right \} }{|\Phi_0(x)|^{3}}, 
\label{eqSGfull2}
\end{align} 
where $\{X,Y\}\equiv X \partial_x Y-(\partial_x X)Y$, and
\begin{equation}
H=-{1 \over 2}{d^2 \over dx^2}
+{q^2 \over 2 |\Phi_0(x)|^{4}}
+V(x)-1+{q^2 \over 2}+|\Phi_0(x)|^{3}.
\label{eq.ap2}
\end{equation} 
The boundary condition in terms of $S(x)$ and $G(x)$ is given by 
\begin{align}
\begin{pmatrix}
S \\ G 
\end{pmatrix}
= &
\begin{pmatrix}
\tilde{S}_{k_{1}} \\ \tilde{G}_{k_{1}}
\end{pmatrix}
e^{i k_{1} x} 
+ 
R
\begin{pmatrix}
\tilde{S}_{k_{2}} \\ \tilde{G}_{k_{2}} 
\end{pmatrix}
e^{i k_{2} x} 
\nonumber
\\
&
+
A
\begin{pmatrix}
\tilde{S}_{k_{3}} \\ \tilde{G}_{k_{3}}
\end{pmatrix}
e^{i k_{3} x} 
( x \ll -1 ), 
\label{eqBoundarySG1}
\\
\begin{pmatrix}
S \\ G 
\end{pmatrix}
= &
T
\begin{pmatrix}
\tilde{S}_{k_{1}} \\ \tilde{G}_{k_{1}} 
\end{pmatrix}
e^{i k_{1} x} 
\qquad \qquad \qquad 
\nonumber
\\
&
+
B 
\begin{pmatrix}
\tilde{S}_{k_{4}} \\ \tilde{G}_{k_{4}} 
\end{pmatrix}
e^{i k_{4} x} (x \gg +1), 
\label{eqBoundarySG2}
\end{align} 
where 
$(\tilde{S}_{k_{}}, \tilde{G}_{k_{}}) = (1+k^{2}/(4c), (E-qk)/(2c))$ for $k=k_{1,2}$, 
and 
$(\tilde{S}_{k_{}}, \tilde{G}_{k_{}}) = ((E-qk)/(2c) , k^{2}/(4c))$ for $k=k_{3,4}$~\cite{SG}. 
In the low energy regime ($E\ll 1$) with small supercurrent momentum $|q|\ll 1$, one may approximate the spin-wave momentum 
to $k_{1} \simeq E/\sqrt{c}$ $(\equiv p)$. 
Then, in expanding coefficients in terms of $p$ and $q$ as 
\begin{align}
\begin{pmatrix}
T \\ R \\ A \\ B
\end{pmatrix}
= 
\sum\limits_{l,l' = 0}^{\infty}
p^{l} q^{l'}
\begin{pmatrix}
T^{(l,l')} \\ R^{(l,l')} \\ A^{(l,l')} \\ B^{(l,l')}
\end{pmatrix}, 
\end{align}
the boundary condition of the lowest order of $p$ and $q$ is 
given by 
\begin{align}
\begin{pmatrix}
S \\ G
\end{pmatrix} 
= &
\begin{pmatrix}
1 + R^{(0,0)} \\ A^{(0,0)} e^{2\sqrt{c} x}
\end{pmatrix} 
( x \ll -1), 
\label{bndrySG00Left}
\\
\begin{pmatrix}
S \\ G
\end{pmatrix} 
= &
\begin{pmatrix}
T^{(0,0)} \\ B^{(0,0)} e^{-2\sqrt{c} x}
\end{pmatrix} 
( x \gg +1). 
\label{bndrySG00Right}
\end{align}

On the other hand, expanding $S$ and $G$ in terms of $p$ and $q$ as 
\begin{align}
S(x) = & \sum\limits_{l,l' = 0}^{\infty} p^{l} q^{l'} S^{(l,l')}(x), 
\label{eq.ap3}
\\ 
G(x) = & \sum\limits_{l,l' = 0}^{\infty} p^{l} q^{l'} G^{(l,l')}(x), 
\label{eq.ap4}
\end{align} 
we obtain equations for the coefficients in Eqs.(\ref{eq.ap3}) and (\ref{eq.ap4}) as,
\begin{align}
{h}_{0} (x) S^{(0,0)} (x) = & 0, 
\label{eq:S00}
\\
\left [  {h}_{0} (x)+ 2 c_{} \gamma^{2}(x) \right ]  G^{(0,0)} (x)= & 0,
\label{eq:G00}
\\
{h}_{0} (x) S^{(1,0)}      (x)           = & \sqrt{c_{}} G^{(0,0)} (x),
\label{eq:S10}
\\
\left [  {h}_{0} (x) + 2 c_{} \gamma^{2}(x) \right ]  G^{(1,0)} (x) = & \sqrt{c_{}} S^{(0,0)} (x) , 
\label{eq:G10}
\\
{h}_{0} (x) S^{(0,1)}   (x)               = & i {  \left \{ \gamma (x), G^{(0,0)} (x) \right \}  \over \gamma^{3} (x)}, 
\label{eq:S01} 
\\
\left [ {h}_{0} (x) + 2 c_{} \gamma^{2}(x) \right ]    G^{(0,1)} (x) = & i {  \left \{ \gamma (x), S^{(0,0)} (x) \right \}  \over \gamma^{3} (x)},  
\label{eq:G01}
\end{align}
where $h_{0}(x) \equiv - (1/2)\partial_{x}^{2} + V_{0} \delta (x) - 1 + \gamma^{2} (x)$. 
\par
Equation (\ref{eq:S00}) has two independent solutions $f_{\rm e}(x) = \gamma (x)$ and 
$f_{\rm o}(x) = x\gamma (x) + {\rm sgn}(x) [ \gamma (x) - \gamma(0)] /\gamma(0) $, 
both of which satisfy the boundary condition at $x = 0$. 
As a result, $S^{(0,0)}(x)=C^{(0,0)} f_{\rm e}(x) + D^{(0,0)} f_{\rm o}(x)$ follows, 
where $C^{(0,0)}$ and $D^{(0,0)}$ are coefficients. 
Taking $|x| \rightarrow \infty$ and considering the boundary conditions (\ref{bndrySG00Left}) and (\ref{bndrySG00Right}), 
we obtain $C^{(0,0)} = T^{(0,0)} = 1 + R^{(0,0)}$ and $D^{(0,0)} = 0$. 
\par
To solve Eq.(\ref{eq:G00}), it is convenient to rewrite this equation in the form
\begin{eqnarray}
(1-\gamma^{2}) \frac{d^{2} G^{(0,0)}}{d\gamma^{2}} 
- 2 \gamma \frac{d G^{(0,0)}}{d \gamma} 
\nonumber
\\
+ \left [ 2 (1 + 2 c_{}) - \frac{4c_{}}{1-\gamma^{2}} \right ] G^{(0,0)} = 0, 
\label{eq:G00B}
\end{eqnarray} 
where we have used the relation $\partial_{x} \gamma (x) = {\rm sgn}(x) [1 - \gamma^{2} (x)]$. The solution of (\ref{eq:G00}) is given by $G^{(0,0)} (x) = a P_{\nu}^{\eta} (\gamma (x) ) + b Q_{\nu}^{\eta} (\gamma (x) )$, where $P_{\nu}^{\eta}$ and $Q_{\nu}^{\eta}$ are the associated Legendre functions~\cite{Legendre}, with $\nu = (-1 + \sqrt{9 + 16 c_{}})/2$ and $\eta = 2 \sqrt{c_{}}$. When $\eta$ is a non-integer, the asymptotic behaviors of the associated Legendre functions are given by
\begin{align}
P_{\nu}^{\eta} (\gamma (x) ) \simeq & \frac{ e^{\eta \pi i} e^{\eta x_{0}} }{\Gamma (1-\eta)} e^{\eta |x|}, 
\\
Q_{\nu}^{\eta} (\gamma (x) ) \simeq & \frac{\pi}{2} \frac{1}{\tan (\eta \pi)} \frac{e^{\eta \pi i} e^{\eta x_{0}} }{\Gamma (1-\eta)} e^{\eta |x|}. 
\end{align}
When we take $b = - a (2/\pi) \tan (\eta \pi) ( \equiv a {\tilde b}) $, the function $G^{(0,0)} (x) = a [ P_{\nu}^{\eta} (\gamma (x) ) + {\tilde b} Q_{\nu}^{\eta} (\gamma (x) )]$ converges. However, imposing the boundary condition at $x = 0$, $\partial_{x} G^{(0,0)} (+ 0) - \partial_{x} G^{(0,0)} (- 0) = 2 V_{0} G^{(0,0)} (0)$, one obtain $a = 0$, leading to $G^{(0,0)} (x) = 0$. When $\eta$ is an integer, while $P_{\nu}^{\eta} (\gamma (x\to\pm\infty))$ converges as $\exp{(-\eta x_{0})} \Gamma(\nu + \eta + 1) / [\eta! \Gamma(\nu - \eta + 1)] \exp{(-\eta|x|)}$, $Q_{\nu}^{\eta} (\gamma (x))$ diverges as $Q_{\nu}^{\eta}\propto \exp{(\eta |x|)}$ ($|x| \gg 1$). Thus, we need to take $b=0$. However, the boundary condition at $x=0$ again gives $G^{(0,0)}(x)=0$. Thus, one concludes $G^{(0,0)} (x) = 0$. 
Considering (\ref{bndrySG00Left}) and (\ref{bndrySG00Right}), we have $A^{(0,0)} = B^{(0,0)} = 0$. 
\par
The boundary conditions (\ref{eqBoundarySG1}) and (\ref{eqBoundarySG2}) are then given by
\begin{align}
\begin{pmatrix}
S \\ G 
\end{pmatrix}
= 
\begin{pmatrix}
T^{(0,0)} + p [R^{(1,0)} + ix (1-R^{(0,0)})] + q R^{(0,1)}
\\
p/(2\sqrt{c}) + q A^{(0,1)} e^{2\sqrt{c}x}
\end{pmatrix}
\nonumber
\\
(x \ll -1)
\label{boundaryL2}
\\
\begin{pmatrix}
S \\ G 
\end{pmatrix}
= 
\begin{pmatrix}
T^{(0,0)} + p (T^{(1,0)} + ix T^{(0,0)}) + q T^{(0,1)}
\\
p/(2\sqrt{c}) + q B^{(0,1)} e^{-2\sqrt{c}x}
\end{pmatrix}
\nonumber
\\ 
(x \gg +1)
\label{boundaryR2}
\end{align} 
within the accuracy of $O (p)$ and $O (q)$, 
where we took $p|x| \ll 1$. 
\par
Noting that $G^{(0,0)} (x) = 0$, one finds that Eqs. (\ref{eq:S10}) and (\ref{eq:S01}) 
are the same as Eq. (\ref{eq:S00}). Thus, $S^{(1,0)}(x)$ and $S^{(0,1)} (x)$ are both given by the linear combination of $f_{\rm e}(x)$ and $f_{\rm o}(x)$. Using this, we may write $S(x)$ in the form, within the accuracy of $O(p)$ and $O(q)$, 
\begin{align}
S (x) = & 
T_{}^{(0,0)} f_{\rm e} (x) + p \left ( C^{(1,0)} f_{\rm e} (x) + D^{(1,0)} f_{\rm o} (x) \right )
\nonumber
\\ & 
+ q \left (  C^{(0,1)} f_{\rm e} (x) + D^{(0,1)} f_{\rm o} (x) \right ). 
\label{eqSL}
\end{align}
$C^{(l,l')}$ and $D^{(l,l')}$ $(l,l' = 0,1$) are coefficients. 
\par
Comparing Eqs. (\ref{eqSL}) for $|x| \gg +1$ with Eqs. (\ref{boundaryL2}) and (\ref{boundaryR2}), 
we have $(T^{(0,0)}, R^{(0,0)}, D^{(1,0)}, D^{(0,1)}) = (1,0,i,0)$, 
$C^{(1,0)} = T^{(1,0)} = R^{(0,1)}$, and $T^{(1,0)} - T^{(0,1)} = - (R^{(1,0)} - R^{(0,1)}) = i [-1 + 1/\gamma (0)]$. 
Using this result, one finds, 
\begin{align}
S (x) =&  [ 1 + i ( p C^{(1,0)} + q C^{(0,1)} ) + i p x ] \gamma (x) 
\nonumber 
\\ 
& + {\rm sgn}(x) i p [ \gamma (x) - \gamma (0) ] /\gamma (0). 
\end{align} 
As a result, we obtain,  within the accuracy of $O(p)$ and $O(q)$, 
\begin{align} 
S (x) =  & e^{i \alpha_{p,q}} \Psi_{<} (x,q) (x < 0), 
\label{eq:61_1}
\\
S (x) =  & e^{i \alpha_{p,q}} \Psi_{>} (x,q) (x \geq 0), 
\label{eq:61_2}
\end{align} 
where $\alpha_{p,q} \equiv p C^{(1,0)} + q C^{(0,1)}$. 
\par
Next, we evaluate $G^{(0,1)}$ and $G^{(1,0)}$. Substituting $S^{(0,0)}=\gamma(x)$ into Eqs. (\ref{eq:G01}) and (\ref{eq:G10}), we obtain
\begin{align}
[h_{0} (x) + 2 c_{} \gamma^{2} (x)] G^{(0,1)} (x) = & 0, 
\label{eq:ReG01}
\\
[h_{0} (x) + 2 c_{} \gamma^{2} (x)] G^{(1,0)} (x) = & \sqrt{c_{}} \gamma(x). 
\label{eq:ReG10}
\end{align} 
As in the case of $G^{(0,0)}$ (which obeys Eq. (\ref{eq:G00})), Eq. (\ref{eq:ReG01}) gives $G^{(0,1)}=0$. 
Considering (\ref{boundaryL2}) and (\ref{boundaryR2}), we have $A^{(0,1)} = B^{(0,1)} =0$. 
For Eq.(\ref{eq:ReG10}), we have, within the accuracy of $O (p)$ and $O (q)$,  
\begin{align}
G (x) = & p [ a^{(1,0)} P_{\nu}^{\eta} (\gamma (x) ) + b^{(1,0)} Q_{\nu}^{\eta} (\gamma (x) ) 
\nonumber
\\ &
+ G_{\rm s}^{(1,0)} (x) ], 
\label{eq:A31}
\end{align}
where 
\begin{align}
G_{\rm s}^{(1,0)} (x) = & 
\left [ 
2 \sqrt{c}  \int_{x_{1}}^{x} \frac{{\rm sgn} (x)}{\Delta} Q_{\nu}^{\eta} (\gamma (x) ) \gamma (x) dx 
\right ] 
P_{\nu}^{\eta} (\gamma (x) )
\nonumber 
\\
& - 
\left [ 
2 \sqrt{c} \int_{x_{2}}^{x} \frac{{\rm sgn} (x)}{\Delta} P_{\nu}^{\eta} (\gamma (x) ) \gamma (x) dx 
\right ] Q_{\nu}^{\eta} (\gamma (x) ). 
\end{align} 
$x_{1}$ and $x_{2}$ are constants and $\Delta$ is defined as~\cite{Wronski} 
\begin{align}
\Delta \equiv 2^{2\eta} e^{2 \eta \pi i} 
\frac{\Gamma ( (\nu + \eta + 1)/2)   \Gamma ( (\nu + \eta)/2 + 1) }
{\Gamma ( (\nu - \eta + 1)/2)   \Gamma ( (\nu - \eta)/2 + 1)}. 
\label{DeltaDef}
\end{align}
Considering the asymptotic behavior of $P_{\nu}^{\eta} (\gamma (x) )$ and $Q_{\nu}^{\eta} (\gamma (x) )$ for $x = \pm\infty$, 
we find that $G_{\rm s}^{(1,0)} (x)$ converges. 

Since $G (x)$ should converge for $x = \pm\infty$ in this tunneling problem, 
we have $b^{(1,0)} = a^{(1,0)} \tilde{b}$ for $\eta$ being a non-integer and $b^{(1,0)} = 0$ for $\eta$ being an integer 
as in the same discussion for $G^{(0,0)} (x)$. 
We still have three unknown parameters $(a^{(1,0)}, x_{1}, x_{2})$, but these three parameters are determined from three boundary conditions: 
First one is the boundary condition at $x=0$ given by $\partial_{x} G(+0) - \partial_{x}G (-0) = 2 V_0 G(0)$. 
Second and third ones are those for $x = \pm\infty$, given by $G^{(1,0)} (x = \pm\infty)= 1/(2\sqrt{c})$ 
where we used (\ref{boundaryL2}) and (\ref{boundaryR2}).

From Eqs. (\ref{ApproxDeltaPhi01}), (\ref{ApproxDeltaPhi02}), (\ref{defS}), (\ref{defG}), (\ref{eq:61_1}), (\ref{eq:61_2}),and (\ref{eq:A31}), we have 
\begin{align}
\left(
\begin{array}{c}
\phi_{+1} \\
\phi^*_{-1}
\end{array}
\right) 
\simeq
\left(
\begin{array}{c}
+e^{i \alpha_{p,q}} \Psi_{<}^{}(X_{-},p+q) /2 \\
- e^{i \alpha_{p,q}} \Psi_{<}^{} (X_{+},p-q) /2
\end{array}
\right) 
, (x < 0), 
\\
\left(
\begin{array}{c}
\phi_{+1} \\
\phi^*_{-1}
\end{array}
\right)
\simeq
\left(
\begin{array}{c}
+e^{i \alpha_{p,q}} \Psi_{>}(X_{+},p+q) /2\\
- e^{i \alpha_{p,q}} \Psi_{>}^{} (X_{-},p-q) /2
\end{array}
\right) 
, (x \geq 0), 
\end{align}
within the accuracy of $O (p)$ and $O  (q)$ 
if we use $X_{\pm} \equiv x \pm G(x)/[1-\gamma^{2}(x)]$. 
Taking $\Lambda_\pm=\pm e^{i\alpha_{p,q} } /2$, we obtain Eqs. (\ref{eq.super}) and (\ref{eq.super2}).





\begin{thebibliography}{99}
\bibitem{NoteZero} We have numerically confirmed that the barrier potential $V(x)$ does not induce the $\Phi_{\pm 1}(x)$-components. Thus, Eq.(\ref{eq.5}) holds everywhere in the present tunneling problem.
\bibitem{Polar} Note that fluctuations of the spin density $\Delta F_{i} \equiv [ \hat{\Phi}^{\dag} S_{i}^{2} \hat{\Phi} - (\hat{\Phi}^{\dag} S_{i} \hat{\Phi})^{2} ]^{1/2}$ for $i = x,y$ do not vanish ($\Delta F_{x,y} = |\Phi_{0}|$) although $\Delta F_{z} = 0$.
\bibitem{note100} The phase $\varphi_0$ behaves as $\varphi_0=qx+C_\pm$ when $x=\pm\infty$. The phase difference is defines as $\Delta\varphi=C_+-C_-$.
\bibitem{note} The sign difference of the off-diagonal components between Eq.(\ref{BogoPM1}) and Eq.(\ref{Bogo}) can be absorbed into the condensate wavefunction by the gauge transformation $\Phi_0(x)\to\Phi_0(x)e^{i\pi/2}$.
\bibitem{SG} $\tilde{S}_{k}$ and $\tilde{G}_{k}$ are determined from $(\ref{eqSGfull1})$ and $(\ref{eqSGfull2})$, 
and we choose $(\tilde{S}_{k}, \tilde{G}_{k})$ so as to begin with $O  (1)$ for small $p$ and $q$.
\bibitem{Legendre} In this paper, we use the associated Legendre functions in Ferrers normalization (e.g., $P_{\nu}^{\eta} (x) \equiv [\exp{(\eta\pi i)}/\Gamma (1-\eta)] [(1+x)/(1-x)]^{(\eta/2)} F (-\nu, \nu + 1; 1- \eta; (1-x)/2)$, where $F(a,b;c;z)$ is the Gauss hyper-geometric function). For $Q_{\nu}^{\eta}(x)$, we use the relation 
$Q_{\nu}^{\eta} (x) = \{\pi/[2\sin(\eta\pi)]\} \{ \cos(\eta\pi) P_{\nu}^{\eta} (x) - [ \Gamma (\nu+ \eta + 1)/\Gamma(\nu-\eta+1)] P_{\nu}^{-\eta} (x) \}$ when $\eta$ is a non-integer.
\bibitem{Wronski} (\ref{DeltaDef}) is determined from the Wronskian for the associated Legendre function given by $P_{\nu}^{\eta} (x) \partial_{x} Q_{\nu}^{\eta} (x) - Q_{\nu}^{\eta} (x) \partial_{x} P_{\nu}^{\eta} (x) = \Delta/(1-x^{2})$.
\end{thebibliography}
\end{document}